\documentclass[twocolumn]{aastex701}
\usepackage{soul}

\shorttitle{Bar Kinematics in Paired and Isolated Galaxies}
\shortauthors{Aziz et al.}

\begin{document}

\title{Bar Kinematics in Paired and Isolated Galaxies: Slow Bars Are More Likely Found in Paired Galaxies}

\author{Brian Aziz}
\affiliation{David A. Dunlap Department of Astronomy \& Astrophysics, University of Toronto, 50 St. George St., Toronto, ON M5S 3H4, Canada}
\email{brian.aziz@mail.utoronto.ca}

\author{Tobias Géron}
\affiliation{Dunlap Institute for Astronomy \& Astrophysics, University of Toronto, 50 St. George Street, Toronto, ON M5S 3H4, Canada}
\email{tobias.geron@utoronto.ca}

\author{Izzy L. Garland}
\affiliation{Department of Theoretical Physics and Astrophysics, Faculty of Science, Masaryk University, Kotlářská 2, Brno, 611 37, Czech
Republic}
\affiliation{School of Physics and Astronomy, Lancaster University, Lancaster LA1 4YB, UK}
\email{garland@mail.muni.cz}

\author{David O'Ryan}
\affiliation{European Space Agency (ESA), European Space Astronomy Centre (ESAC), Camino Bajo del Castillo s/n, 28692, Villanueva de
la Cañada, Madrid, Spain}
\email{david.oryan@esa.int}

\author{Hayley Roberts}
\affiliation{Minnesota Institute for Astrophysics, University of Minnesota, 116 Church Street SE, Minneapolis, MN 55455, USA}
\affiliation{School of Physics and Astronomy, University of Minnesota, 116 Church Street SE, Minneapolis, MN 55455, USA}
\email{robertsh@umn.edu}

\author{Rebecca J. Smethurst}
\affiliation{Oxford Astrophysics, Department of Physics, University of Oxford, Denys Wilkinson Building, Keble Road, Oxford OX1 3RH, UK}
\email{rebecca.smethurst@physics.ox.ac.uk}

\begin{abstract}
    Bars are common structures in disk galaxies that can strongly influence the evolution of their hosts. Simulations predict that bars formed through tidal interactions tend to rotate slower than bars formed in isolation. We test this prediction observationally by studying the kinematics of 314 bars in galaxies between $0.016<z<0.130$. We use estimates of the bar pattern speed ($\Omega_p$) and the rotation rate ($\mathcal{R}$) obtained from four studies using IFU data from the MaNGA survey. We cross-match the host galaxies of the bars to the Galaxy Zoo DESI catalogue in order to study their environment and classify each galaxy as paired or isolated. We find that bars that are slower in terms of $\mathcal{R}$ are more likely to be found in paired galaxies. The fraction of bars in paired galaxies increases from $5^{+6}_{-2}\%$ for $0<\mathcal{R}<1.0$ to $31^{+15}_{-9}\%$ for $3.0<\mathcal{R}<4.0$.  Our results also suggest a possible difference in the distribution of $\mathcal{R}$ between paired (mean $\mathcal{R}=2.1 \pm 0.2$) and isolated (mean $\mathcal{R}=1.75 \pm 0.07$) galaxies (p-value=0.037; 2.09$\sigma$). On the other hand, our analysis reveals no noticeable trend between the fraction of paired galaxies and the bar pattern speed, which indicates that the pattern speed is not correlated with the likelihood of a galaxy being paired. We also find no difference in the distribution of $\Omega_p$ between paired and isolated galaxies (p-value$>$0.250; $<$1.15$\sigma$).
\end{abstract}


\section{Introduction}\label{sec:intro}

Galaxy bars are central, dynamical structures found in roughly 30-60\% of disk galaxies in the local Universe \citep{marinova2007,aguerri2009,masters2011,geron2021}. Bars remain common at higher redshifts, as bar fractions of roughly 10-20\% have been measured at $z=2$ \citep{LeConte2024,geron2025}. The likelihood of a disk hosting a bar strongly depends on the colour and  stellar mass of its host galaxy \citep{aguerri2009,nair2010,mendez-abreu2012,erwin2018}. Moreover, bars can strongly influence the dynamics and evolution of their host galaxies. They are responsible for the redistribution of angular momentum between different components of the galaxy, such as the disk, the galactic bulge, and the dark matter halo \citep{lynden-bell1978,debattista1998,athanassoula2003,martinez-valpuesta2006,romeo2023}. Bars also funnel gas towards the central regions of a galaxy \citep{athanassoula1992,sakamoto1999,fragkoudi2016}, potentially triggering bursts of star formation \citep{mihos1994,sheth2005}. Moreover, bars have been linked to the quenching of their host galaxies, as they appear more often in quiescent galaxies \citep{masters2012,cheung2013,kruk2018}. In short, bars are remarkably prevalent structures that highly influence the evolution of their host galaxies.

The extent to which a galaxy is affected by its bar is dependent, in part, on the kinematics of the bar (see \citealp{geron2024}). The kinematics of a bar can be described using the bar pattern speed ($\Omega_p$), which is defined as the rotational frequency of the bar about the galactic centre \citep{athanassoula2003}. It is one of the most important dynamical parameters of a bar as it helps to understand how barred galaxies evolve. For instance, the bar pattern speed is linked to the rate of angular momentum exchange between the bar and its host galaxy \citep{debattista2000,athanassoula2003,geron2026}. Several methods have been proposed to measure the bar pattern speed. Some works have utilized hydrodynamical simulations of galaxies to match the modeled and observed gas velocity fields and recover $\Omega_p$ \citep{hunter1988,lindblad1996,rautiainen2008}. Other methods use the location of prominent morphological features such as spiral arms \citep{puerari1997,aguerri1998} or galaxy rings \citep{rautiainen2000,perez2012} to derive the pattern speed. However, methods such as these are model-dependent. The only model-independent method to date is the Tremaine-Weinberg (TW) method \citep{TW1984}. This method measures the pattern speed using a tracer (i.e. stars or gas) that satisfies the continuity equation. The TW method was originally applied to long-slit spectroscopic observations of early-type barred galaxies \citep{Merrifield1995,debattista2002,Debattista2004,Corsini2007}. With the development of integral field spectroscopy, the TW method has been applied to a growing number of galaxies \citep{Aguerri2015,Cuomo2019,Williams2021}. Most notably, \citet{geron2023} accurately measured the pattern speed of 225 bars from the Mapping Nearby Galaxies at Apache Point Observatory (MaNGA) survey, and it is the largest such sample to date.

The bar pattern speed also defines the corotation radius $R_{\rm CR}$. This is the radius at which stars in the disk have the same angular rotation rate as the bar. If $R_{\rm CR}$ and the bar radius $R_{\rm bar}$ are known, the dimensionless ratio $\mathcal{R}=R_{\rm CR}/R_{\rm bar}$ can be calculated. This is known as the bar rotation rate, and it is used to classify bars as `fast' ($\mathcal{R}<1.4$) and `slow' ($\mathcal{R}>1.4$) \citep{debattista2000,rautiainen2008}. It is suggested that bars cannot extend beyond their corotation radius \citep{contopoulos1980,athanassoula1992}, which means that values of $\mathcal{R}<1.0$ are nonphysical. However, bars with such values, termed `ultrafast' bars, have been repeatedly measured in observational work \citep{Guo,garma-oehmichen2020,garma-oehmichen2022,geron2023}. This is suspected to be due to incorrect measurements of $R_{\rm bar}$ \citep{cuomo2021,roshan2021,frankel2022}. In any case, the bar rotation rate $\mathcal{R}$ is an important parameter as it provides a distinction between fast and slow bars.

Bars have two most commonly studied formation mechanisms. They can form in isolation, after being triggered by gravitational instabilities within the disk of the galaxy \citep{hohl1971,ostriker1973,sellwood2014,lokas2019}. Alternatively, bars can be induced by tidal interactions during flybys, collisions, or mergers \citep{Byrd1986,gauthier2006,noguchi1987,lokas2016,lokas2019}. In general, galaxy interactions play a fundamental role in galaxy evolution. They are capable of increasing star formation rates (SFRs; \citealp{barnes1991,berrier2006,scott2014,knapen2016}), facilitating quenching \citep{hopkins2013,das2023}, and potentially igniting active galactic nuclei (AGNs; \citealp{ellison2011,comerford2015,hernandez-toledo}). In the context of bar formation, however, galaxy interactions and mergers can influence the dynamics of a galaxy to directly trigger the formation of a bar. The dynamic state of the disk matters as well. In cold and warm disks, simulations suggest that bar formation in dense environments can be due to a combination of both internal instabilities of the disk and tidal interactions \citep{zheng&shen1}. However, simulations have shown that in kinematically hotter disks, tidal interactions alone can trigger bar formation \citep{martinez-valpuesta-2017,zheng&shen1}. Other environmental phenomena have been suggested to form bars, such as cluster-cluster interactions \citep{Yoon2019,Yoon2020}. However, subsequent work with a larger sample by \citet{Deng2023} has shown that this evidence is in fact weaker than previously reported.

Simulations have predicted a correlation between the formation mechanism of the bar and its kinematics: bars formed through tidal interactions are typically slow, whereas bars formed in isolation are typically fast. Simulations typically focus on the behavior of $\mathcal{R}$ in tidally induced bars, and find values of $\mathcal{R}>1.4$ \citep{lokas2014,lokas2016,gajda2017,gajda2018}. Moreover, some simulations show that bars formed through tidal interactions in kinematically hot disks (where bars cannot form in isolation) are initially slower, in terms of both $\mathcal{R}$ and $\Omega_p$, than bars formed in isolation in cold and warm disks \citep{martinez-valpuesta2016,martinez-valpuesta-2017,zheng&shen1}. These simulations find that, over time, the $\mathcal{R}$ values of these tidally induced bars decrease and eventually enter the fast regime after about 4 Gyr. However, for bars formed in isolation, $\mathcal{R}$ fluctuates but remains less than $\mathcal{R}=1.4$ for most of the time. Moreover, the bar pattern speed remains constant and relatively small (10-20 km s$^{-1}$ kpc$^{-1}$) in tidally induced bars in kinematically hot disks. However, for bars formed in isolation, $\Omega_p$ decreases with time to eventually reach a range similar to that of the tidally induced bars \citep{martinez-valpuesta-2017,zheng&shen1}. Nevertheless, initially, the tidally induced bars are slower in terms of both $\mathcal{R}$ and $\Omega_p$. In short, simulations suggest that bars formed through tidal interactions tend to be slower than those formed by internal instabilities in isolation.

In previous work, the prediction made by the simulations has been tested by investigating the kinematics of bars in galaxies in high-density and low-density environments (see \citealp{puczek2026}). It was found that galaxies in high-density environments tend to host bars that are slower in $\mathcal{R}$. Moreover, \citet{Cuomo2022} observed a bar in a dwarf galaxy belonging to an infalling group where tidal interactions or flybys are possible. The observed bar was slow and its morphology was consistent with tidal interactions, which led to the suggestion that the bar may have been tidally induced. However, the systematic effect of direct galaxy interactions on bar kinematics remains largely unexplored observationally.

In this work, we provide an observational test of the prediction made by the simulations by comparing the kinematics of bars in paired and isolated galaxies. We compile a catalogue of 314 barred galaxies between $0.016<z<0.130$ from four studies \citep{Guo,garma-oehmichen2020,garma-oehmichen2022,geron2023}. These studies obtained measurements of the bar pattern speed $\Omega_p$ and rotation rate $\mathcal{R}$ for their bars using Integral Field Unit (IFU) data from the MaNGA survey. We cross-match the host galaxies of the bars to the Galaxy Zoo Dark Energy Spectroscopic Survey (GZ DESI) catalogue \citep{GZDESI}, where we inspect the local environment of the galaxies, meaning the nearby galaxies in the immediate surroundings of each barred galaxy. Through inspection of the local environment, we classify each barred galaxy as either paired or isolated. We then test the prediction that bars formed through tidal interactions tend to be slower than those formed through internal instabilities in isolation.

The structure of this paper is as follows. In Section \ref{sec:data&methods}, we introduce our sample and describe how the measurements of $\Omega_p$ and $\mathcal{R}$ were obtained. We also introduce our method of classifying galaxies as paired or isolated, and how we distinguish between major and minor interactions. In Section \ref{sec:results}, we present our results, which are then discussed in relation to simulations in Section \ref{sec:discussion}. Finally, we summarize our main findings in Section \ref{sec:conc}. We assume a standard flat $\Lambda$CDM cosmological model with $H_{\rm 0}=$ 70 km s$^{-1}$ Mpc$^{-1}$, $\Omega_{\rm m}=$ 0.3, and $\Omega_{\Lambda}=$ 0.7 where necessary. We implemented calculations involving the assumed cosmological model with \texttt{Astropy} \citep{astropy2013,astropy2018,astropy2022}.

\section{Data and Methods}\label{sec:data&methods}

\subsection{Data}\label{subsec:data}

This work makes use of data products combined from three sources: the DESI Legacy Imaging Surveys (DESI LS), the GZ DESI catalogue, and the MaNGA survey. We use the GZ DESI catalogue, introduced in Section \ref{subsubsec:GZ_DESI_data}, to inspect the local environment of the barred galaxies in our sample. Throughout this work, we use `local environment' to refer specifically to nearby galaxies in the immediate surroundings of the barred galaxies. We use the local environment to identify potential companions and classify each barred galaxy as either paired or isolated. In the GZ DESI catalogue, images of the barred galaxies were sourced from the DESI LS, which is outlined in Section \ref{subsubsec:DESI_data}. Moreover, the kinematic measurements of the bars adopted in this work were obtained in previous studies using IFU data from the MaNGA survey, described in Section \ref{subsubsec:MaNGA_data}.

\subsubsection{The DESI Legacy Imaging Surveys}\label{subsubsec:DESI_data}

The DESI LS \citep{Dey2019} are a combination of three projects: the Dark Energy Camera Legacy Survey (DECaLS), the Beijing Arizona Sky Survey (BASS), and the Mayall $z$-band Legacy Survey (MzLS), which together cover $\sim 14,000$ deg$^2$ of extragalactic sky in three optical bands ($g$, $r$, and $z$). The uniformity and wide sky coverage of the DESI LS make it optimal for large-scale studies of galaxies and galaxy structures. In particular, the imaging of the DESI LS is relatively deep, as the point-spread function depth of the $r$-band is 23.54 for DECaLS \citep{Dey2019}. This depth is sufficient to resolve morphological features in nearby galaxies, including bars, in varying environments \citep{Walmsley2022,GZDESI}. As such, the GZ DESI catalogue, which is used in this work to investigate the local environment of our barred galaxies, sources images of galaxies from the DESI LS.

\begin{table*}
    \centering
\begin{tabular}{lcc}
\hline
Study & Number of $\Omega_p$ Measurements& Number of $\mathcal{R}$ Measurements\\
\hline
\citet{Guo}& 27 & 27 \\
\citet{garma-oehmichen2020} & 5 & 5 \\
\citet{garma-oehmichen2022} & 58 & 58 \\
\citet{geron2023} & 224 & 210 \\
\hline
\textbf{Total} & \textbf{314} & \textbf{300} \\
\hline
\end{tabular}
\caption{Number of $\Omega_p$ and $\mathcal{R}$ measurements from each study in our sample.}
\label{tab:studies}
\end{table*}

\subsubsection{The Galaxy Zoo DESI Catalogue}\label{subsubsec:GZ_DESI_data}

The GZ DESI catalogue \citep{GZDESI} contains detailed morphology measurements for 8.67 million galaxies in the DESI LS, with a sky coverage that achieves full overlap with the MaNGA survey, among other surveys. This makes cross-matching our galaxies to the GZ DESI catalogue ideal for the purpose of examining the local environment of our galaxies and identifying potential companions. \citet{GZDESI} selects all sources from the DESI LS with $r$-band magnitude brighter than 19, a threshold selected by visual inspection, beyond which galaxies typically lack meaningful resolved visual morphology. Moreover, to remove stellar contamination, the catalogue selects sources with surface brightness $\mu > 18$ mag arcsec$^{-2}$. Note that we only use the GZ DESI catalogue for its sample selection, described below, and not its morphological classifications.

\citet{GZDESI} cross-matched their morphology catalogue to several external catalogues to obtain redshift estimates for their galaxies. They obtain Sloan Digital Sky Survey (SDSS) spectroscopic redshifts from the NASA-Sloan Atlas (NSA, \citealp{z_NSA}), along with spectroscopic and photometric redshifts from the DESI photometric redshift catalogue by \citet{Zhou2021}. We combine these redshifts with the redshift catalogue of DESI Data Release 1 (DR1; \citealp{desicollaboration2026datarelease1dark}). The galaxies in our sample were matched to the GZ DESI catalogue based on their coordinates, with a threshold of 10 arcsec. We then select redshift measurements in the following order of priority: spectroscopic redshifts from DESI DR1, the \texttt{spec\_z} column (spectroscopic redshifts) from \citet{Zhou2021}, SDSS spectroscopic redshifts from the NSA, and the \texttt{photo\_z} column (photometric redshifts) from \citet{Zhou2021}. The redshift range of our sample is $0.016 < z < 0.130$.

\subsubsection{The MaNGA Survey}\label{subsubsec:MaNGA_data}

All four studies whose kinematic measurements we adopt \citep{Guo,garma-oehmichen2020,garma-oehmichen2022,geron2023} obtained resolved stellar velocity and stellar flux data from the MaNGA survey \citep{bundy2015}. This data is necessary in order to apply the TW method, which is used to measure the bar pattern speed, as detailed in Section \ref{subsubsec:kin_methods}. The MaNGA survey is one of three major programs of the SDSS-IV international collaboration \citep{blanton2017}. It is the largest galaxy survey using the Integral Field Spectroscopy (IFS) technique, having observed over 10,000 galaxies in the local Universe spanning all environments and morphological types. The survey's observations are carried out using the 2.5m Sloan Telescope at Apache Point Observatory \citep{gunn2006}. It uses the Baryon Oscillation Spectroscopic Survey (BOSS), which provides a spectral resolution of $R \sim 2000$ in the wavelength range 3600 $\rm{\AA}$ - 10300 $\rm{\AA}$ \citep{smee2013}. The IFUs of the MaNGA survey consist of 19-127 hexagonally stacked fibres which obtain spectra for a galaxy at multiple positions (see \citealp{drory2015} for more instrumentation details). The use of IFU data has allowed the application of the TW method to increasingly large samples, thus enabling accurate measurements of the bar pattern speed. While this work does not use data from MaNGA directly, all measurements of the bar pattern speed and rotation rate used in this work were obtained using IFU data from the survey, as detailed in Section \ref{subsubsec:kin_methods}.

\subsection{Methods}\label{subsec:methods}

We describe the methods with which the four studies \citep{Guo,garma-oehmichen2020,garma-oehmichen2022,geron2023} obtained the kinematic measurements used in our work in Section \ref{subsubsec:kin_methods}. Moreover, we explain the process of classifying each barred galaxy as paired or isolated in Section \ref{subsubsec:pairs_methods}, and the methods used to quantify the strength of the interactions of our paired galaxies in Section \ref{subsubsec:maj/min_methods}.

\subsubsection{Pattern Speed $\Omega_p$ and Bar Rotation Rate $\mathcal{R}$}\label{subsubsec:kin_methods}

We compile a sample of 314 bars with measurements of the bar pattern speed $\Omega_p$ and measurements of the bar rotation rate $\mathcal{R}$. The bar pattern speed is defined as the rotational frequency of the bar about the centre of its host galaxy. On the other hand, $\mathcal{R}=R_{\rm CR}/R_{\rm bar}$ is the dimensionless ratio of the corotation radius to the bar radius, and it is used to classify bars as fast ($\mathcal{R}<1.4$) and slow ($\mathcal{R}>1.4$). These two parameters imply that a bar may be considered slow in terms of either $\Omega_p$ or $\mathcal{R}$, or both. In this work, we extract measurements of bar kinematics from four studies \citep{Guo, garma-oehmichen2020, garma-oehmichen2022, geron2023}, all of which obtained measurements of $\Omega_p$ and $\mathcal{R}$ using IFU data from the MaNGA survey.

Measurements of the bar pattern speed were obtained for all 314 bars using the TW method \citep{TW1984}. This is a model-independent method which measures the pattern speed using a tracer (i.e. stars or gas) that satisfies the continuity equation. The TW method relates the bar pattern speed to the ratio of the luminosity-weighted line-of-sight velocity (also known as the kinematic integral) to the luminosity-weighted position (also known as the photometric integral), while accounting for the inclination of the galaxy. For detailed implementations of the method using IFU observations, see \citet{Guo,garma-oehmichen2020,garma-oehmichen2022,geron2023}.

The reliability of the TW method is limited by certain disk geometries or irregular kinematics. Measurements can become unreliable for disks with very low or very high inclination, or when the bar lies close to the disk's major or minor axis. Moreover, poor stellar velocity maps in galaxies with disturbed structures can limit the TW method's reliability. Appropriate measures were taken to mitigate these limitations by all but one of the works whose kinematic measurements we adopt. \citet{Guo,garma-oehmichen2022,geron2023} place cuts on the inclination and the position angle (PA) difference between the disk and the bar. Moreover, \citet{Guo} removed galaxies with bad velocity dispersion maps or low quality velocity maps, typically from galaxies with disturbed structures or low IFU coverage. \citet{garma-oehmichen2022} removed galaxies that do not exhibit a linear relation between the two TW integrals (the kinematic and photometric integrals), stating that this can occur in systems lacking sufficient symmetry. Moreover, \citet{geron2023} explicitly removed all galaxies with irregular stellar kinematics after inspecting their stellar velocity fields. The only exception is \citet{garma-oehmichen2020}. This work purposely kept unfavourable and disturbed systems in their sample, as one of their goals was to study the uncertainties that arise in the pattern speed from these measurements. Therefore, our sample may contain larger systematic uncertainties in measurements of $\Omega_p$ taken from this work. However, this work only contributes 5 measurements to our sample, which is a relatively small number. As such, the conclusions of this work do not change if we explicitly exclude the 5 galaxies from \citet{garma-oehmichen2020}.

In order to measure $\mathcal{R}=R_{\rm CR}/R_{\rm bar}$, one must first obtain the corotation radius $R_{\rm CR}$ and the bar radius $R_{\rm bar}$. Each of the four studies measured these quantities using different methods. \citet{geron2023} measured the corotation radius by combining their pattern speed measurements with galaxy rotation curves obtained from stellar velocity data from MaNGA (as similarly done in \citealp{garma-oehmichen2020, garma-oehmichen2022}). The authors multiplied the bar pattern speed by a radius range, which quantifies how fast the bar moves as a function of radius. The radius at which this curve intersects the galaxy rotation curve is the corotation radius. \citet{Guo} used the Jeans Anisotropic modeling (JAM) method \citep{cappellari2008} to estimate the disk circular velocity $V_c$, which in turn is used to find the corotation radius by assuming $R_{\rm CR} \approx V_c/\Omega_p$.

The denominator of $\mathcal{R}$, the bar radius, was measured manually by \citet{geron2023}. On the other hand, \citet{garma-oehmichen2020} used the isophotal ellipticity and disk PA profiles to obtain lower and upper limits of the bar radius, which were then used to model the bar radius with a uniform distribution. \citet{garma-oehmichen2022} used the same methods but chose to model the bar radius with a log-normal distribution, this time using the isophotal ellipticity to obtain a mean, and the PA profile to obtain an upper limit. The methods involving isophotal ellipticity and disk PA profiles are described in detail in the corresponding works. Finally, \citet{Guo} measured the bar radius using three methods: the ellipticity radial profile, the PA radial profile and the Fourier decomposition. The final bar radius is taken as the average of the resulting three measurements, and the methods are described in more detail in \citet{Guo}.

\citet{Guo, garma-oehmichen2020, garma-oehmichen2022} obtained measurements of $\mathcal{R}$ for all bars included in our sample. However, of the 224 bars in our sample extracted from \citet{geron2023}, 210 $\mathcal{R}$ measurements are available. In case of duplicates between studies, we always choose the most recent measurement, or the study with an available measurement of $\mathcal{R}$. While our sample includes several `ultrafast' bars ($\mathcal{R}<1.0$), all bars with $\mathcal{R}<1.4$ are referred to as `fast'. The number of measurements taken from each study is shown in Table \ref{tab:studies}.

\subsubsection{Finding Paired Galaxies}\label{subsubsec:pairs_methods}

After cross-matching our sample to the GZ DESI catalogue, we begin the process of classifying galaxies as paired or isolated. For each galaxy in our sample, we first locate all of its neighbouring galaxies within 200 arcsec (this is a projected distance of 278 kpc at the mean redshift of our sample, $z=0.073$). We then select redshifts for the neighbouring galaxies according to the priority order described in Section \ref{subsubsec:GZ_DESI_data}. This is a purposefully large range that ensures we capture any neighbouring galaxy. If no neighbouring galaxies are found within 200 arcsec, our galaxy is automatically classified as isolated.

When a neighbouring galaxy is found, we place thresholds on the projected separation and the rest-frame velocity difference between our galaxy and its tentative companion to determine whether the two truly form an interacting pair. For the projected separation, this threshold is 50 kpc. The velocity difference is calculated with:

\begin{equation}\label{eqn:delta_v}
    \Delta v = \frac{c \times |z_1 -z_2 |}{1+z_{\rm{mean}}}
\end{equation}
where $z_1$ and $z_2$ are the redshifts of the two galaxies, and $z_{\rm{mean}}$ is their average redshift \citep{Duan2025}. The value of the threshold placed on the velocity difference varies according to the source from which the redshift is selected. If both galaxies in our tentative pair have spectroscopic redshifts, then we place a threshold of 1000 km/s on the velocity difference. However, if either our galaxy or its companion is assigned a photometric redshift, we use a threshold of 2500 km/s instead. This is to account for higher uncertainties associated with photometric redshifts. If multiple neighbouring galaxies satisfy both the projected separation and the velocity difference thresholds, as was the case for 3 galaxies in our sample, then the one with the lowest projected separation is selected as the true companion, and the two galaxies form an interacting pair. Otherwise, a galaxy that does not have a pair that satisfies both of these thresholds is classified as isolated. This includes galaxies with companions that satisfy only one of the two thresholds. Our values of the projected separation and rest-frame velocity difference thresholds were chosen in reference to previous studies on merger pairs and interacting galaxies, namely \citet{violino2018, Shah2022, Duan2025}. Our values were guided by those adopted in these studies while taking into account differences in redshift range and in the types of redshift measurements.

\begin{figure*}[ht!]
\plotone{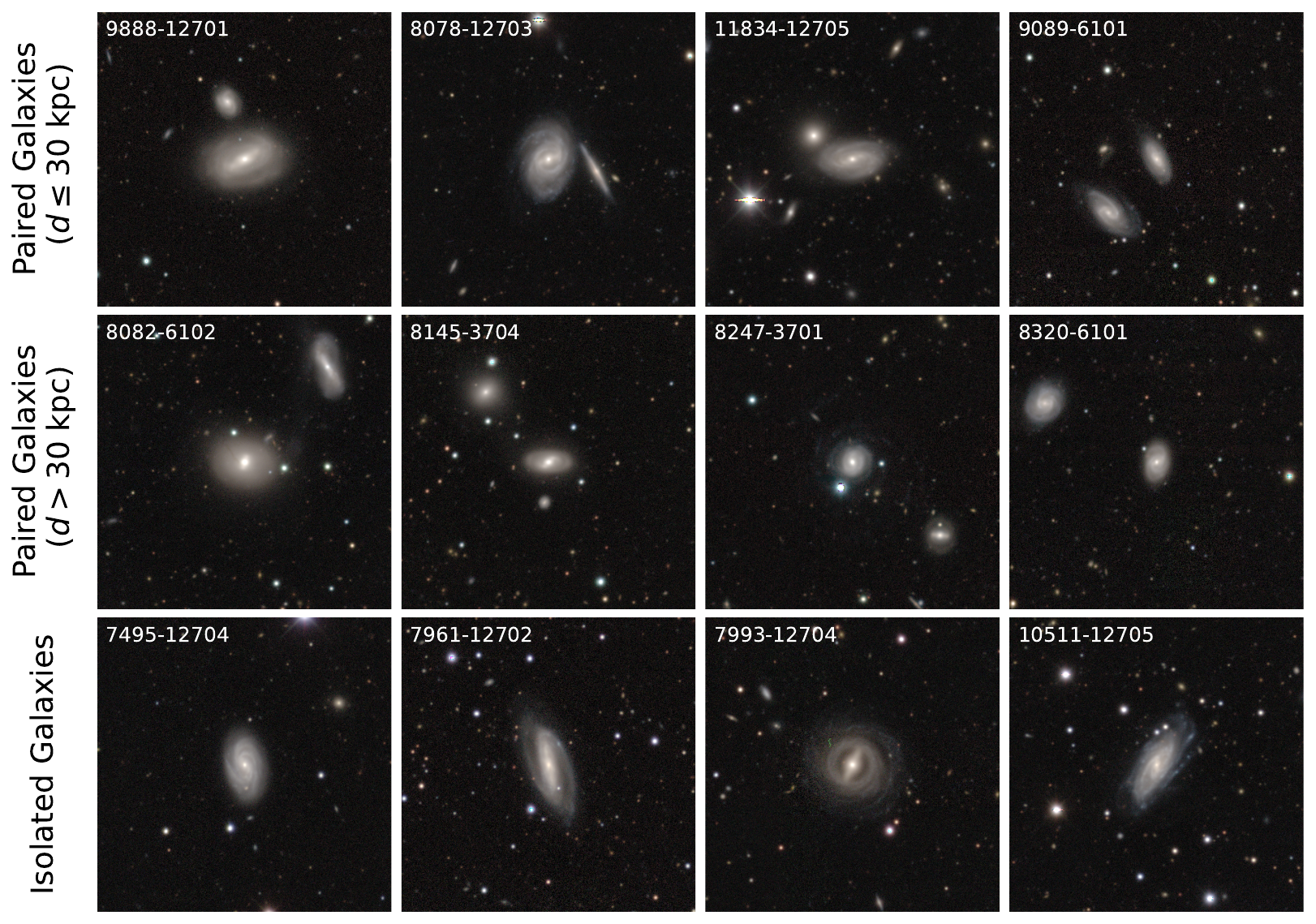}
    \caption{Galaxies in our sample classified as paired with projected distance to the closest companion $d \leq 30$ kpc (top row), paired with $d\ \textgreater\ 30$ kpc (middle row), and isolated (bottom row). The MaNGA plate-IFU number of the galaxy is shown in the top left corner of each image. All cutouts are 212 by 212 arcsec in size and are taken from the DESI Legacy Imaging Surveys.}
    \label{fig:galaxies}
\end{figure*}

This process results in 49 pairs and 265 isolated galaxies. The contrast in the kinematics of the bars between these two populations will be strongest if all isolated bars formed through internal instabilities, while all bars in paired galaxies formed through tidal interactions. However, this will not necessarily hold for all bars in our sample, which is capable of diluting our results. Examples of paired and isolated galaxies are shown in Figure \ref{fig:galaxies}, where we also show that pairs with a projected separation less than 30 kpc typically exhibit clear visual companions. All galaxies have accurate measurements of their bar pattern speeds, while we have 47 paired and 253 isolated galaxies with accurate $\mathcal{R}$ measurements. 

\subsubsection{Quantifying the Strength of the Interaction}\label{subsubsec:maj/min_methods}

To quantify the strength of the interactions involving the paired galaxies in our sample, we use two quantities: stellar mass ratios and magnitude differences. The strength of galaxy interactions and mergers is commonly determined using stellar mass ratios. This method has been found to give significant physical insight into the effect of an interaction on the galaxies involved \citep[e.g., see][]{hopkins2010}. However, stellar mass measurements are not available for all paired galaxies in our sample. Therefore, we apply an additional method which utilizes the absolute $r$-band magnitude difference between a paired galaxy and its companion. This is an observational proxy into the strength of an interaction as opposed to a physical one, but it allows for the full paired galaxy sample to be included in our analysis. We apply both methods independently to the paired galaxies in our sample, and we distinguish major from minor interactions. This will be used to compare the bar kinematics between isolated galaxies, pairs in minor interactions, and pairs in major interactions, as seen in Section \ref{subsec:maj/min_results}. We extract stellar masses for paired galaxies in our sample and their companions from \citet{Zou2019}, and $r$-band magnitudes from DR8 of the DESI LS \citep{Dey2019}.

The \citet{Zou2019} catalogue contains photometric redshifts and stellar masses for galaxies in the DESI imaging survey with $r < 23$ mag. The catalogue covers the redshift range of $z < 1$ and mass range of $8.4 < \log(M_{\textasteriskcentered}) < 11.9$. The stellar masses are obtained using stellar population synthesis model fitting based on photometric redshifts (see Section 4.1 of \citealp{Zou2019}). Out of our 49 pairs, stellar masses were found for both galaxies of 41 pairs. The mass ratio was then obtained as $q_M=M_1/M_2$, where $M_1$ is always the smaller mass of the two in order to ensure $0< q_M < 1$. An interacting pair with $q_M >1/3$ is classified as a major interaction \citep{Conselice2006, French2020}. Using this threshold results in 7 galaxy pairs with major interactions and 34 with minor interactions, out of the 41 pairs with available stellar mass ratios.

Magnitudes were extracted for all 49 paired galaxies and their companions, and the absolute difference between magnitudes in pairs was used to distinguish between major and minor interactions. This is defined as $\Delta m=|m_1 - m_2|$, where $m_1$ and $m_2$ are the magnitudes of the paired galaxy and its companion. An interacting pair with $\Delta m <1.5$ is classified as a major interaction \citep{deRavel2009,Lopez-Sanjuan2015} Using this threshold results in 17 major and 32 minor interactions.

It is important to note that the separate contributions of galaxies in very close or overlapping systems can be difficult to distinguish when extracting stellar masses and magnitudes. The DESI LS measures photometry in the optical bands using \textit{The Tractor} algorithm, which models the position, morphology, and flux of individual sources directly from the imaging data \citep{Dey2019}. This is relevant to the magnitudes used in our work, as this means nearby galaxies are ideally modeled as separate sources and correctly resolved. This is also relevant to the stellar masses we adopt from \citet{Zou2019}, which are derived using the DESI LS photometry. However, \citet{Dey2019} note that large extended sources with complex morphologies can still be poorly modeled by \textit{The Tractor}. Therefore, although the DESI LS reduces blending between nearby sources, sufficiently close or disturbed interacting galaxies may prove to be difficult when it comes to separating the light associated with each galaxy. This could affect both the magnitudes and, in turn, the photometry used to derive the stellar masses in some pairs. However, we believe this is unlikely to affect the galaxies in our sample. This is because, as mentioned in Section \ref{subsubsec:kin_methods}, the TW method cannot be used on highly disturbed and interacting galaxies. Thus, such disturbed galaxies, for which deblending would be an issue, should have been removed from the sample. The top panel of Figure \ref{fig:galaxies} represents the typical close-distance pair in our sample, where even the closest galaxies are still clearly separated.

\section{Results}\label{sec:results}

\subsection{Bar Kinematics of Paired and Isolated Galaxies}\label{subsec:kin_results}

We compare the bar pattern speed $\Omega_p$ and rotation rate $\mathcal{R}$ between paired and isolated galaxies. We first examine how the paired fraction of our sample varies with the kinematic parameters, followed by a direct comparison of the distributions of $\Omega_p$ and $\mathcal{R}$ for paired and isolated galaxies.

\begin{figure*}[ht!]
\includegraphics[width=0.49\textwidth]{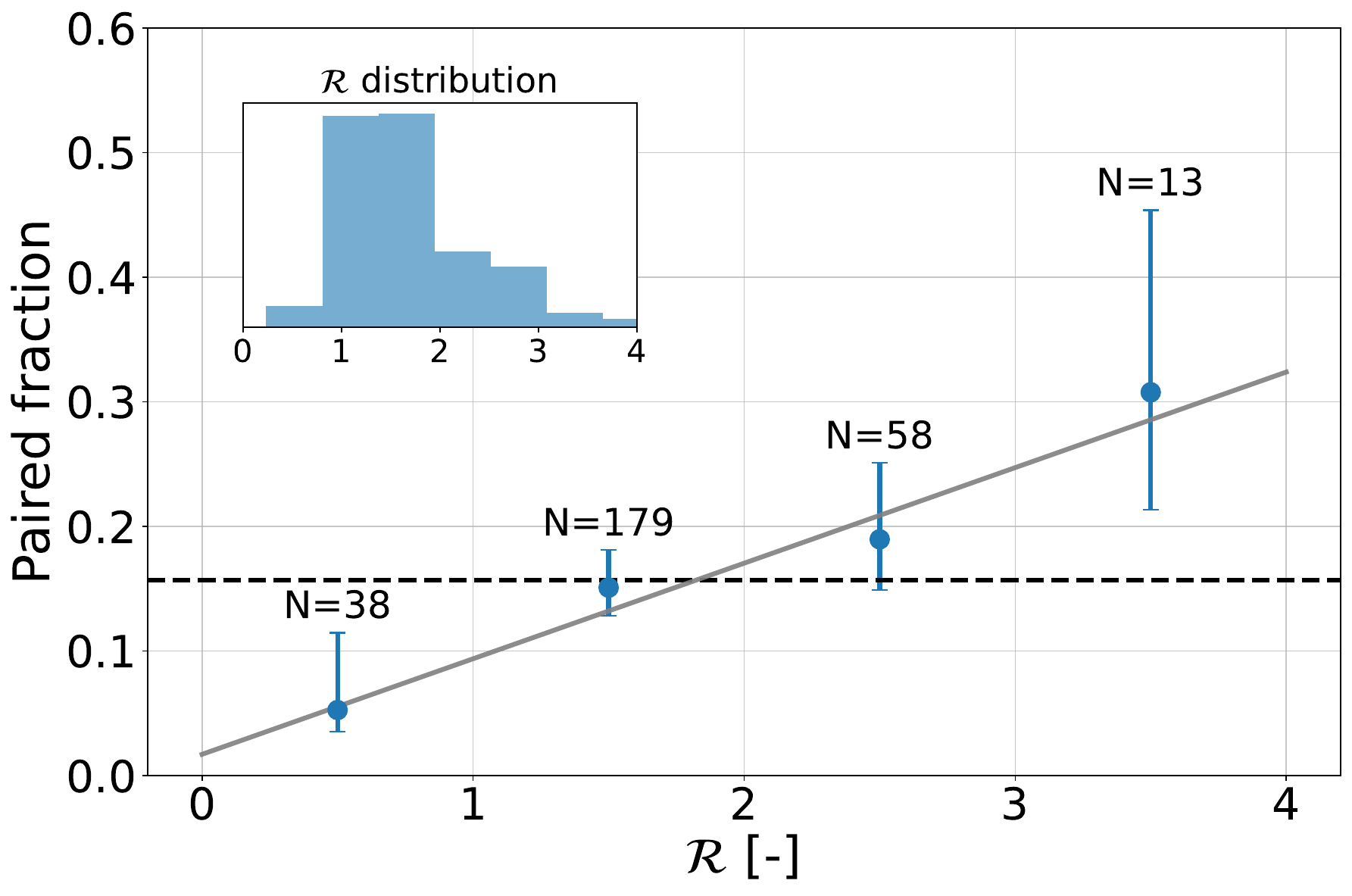}
\includegraphics[width=0.5\textwidth]{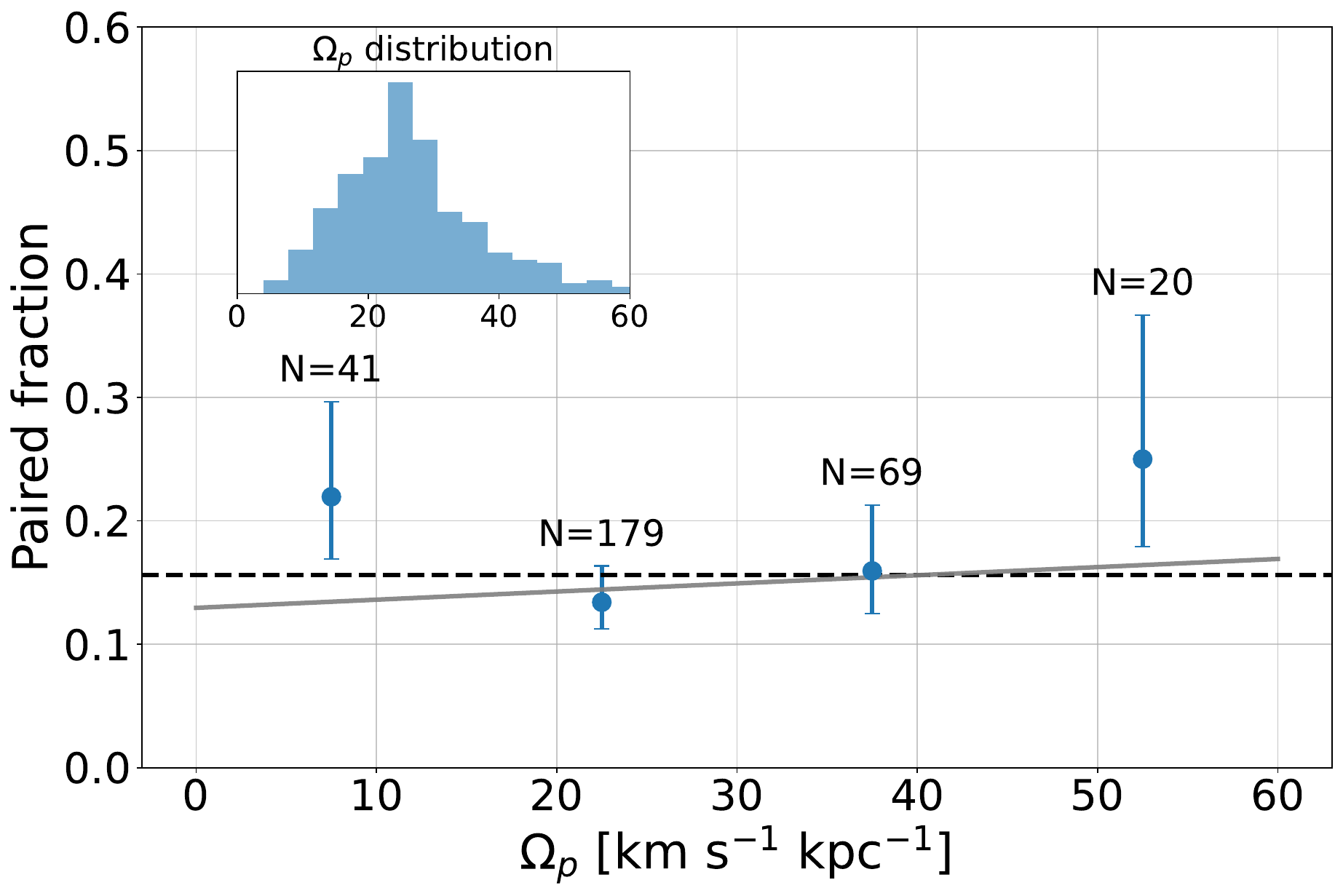}
\caption{The paired fraction of our sample plotted against $\mathcal{R}$ (left) and the bar pattern speed $\Omega_p$ (right), with uncertainties computed using the beta distribution quantile technique \citep{Cameron2011}. A linear fit accounting for asymmetric error is shown to represent the observed trends. An overall increasing trend is seen with  $\mathcal{R}$, where the fit gives a slope of $0.08 \pm 0.02$. On the other hand, no trend is seen with $\Omega_p$ as the slope is $0.0007 \pm 0.002$. We also show the distributions of $\mathcal{R}$ and $\Omega_p$ in the top left corners of each plot, as well as the number of measurements included in each bin. The horizontal dashed line in each plot indicates the paired fraction of the entire sample, which is equal to 16\%.}
\label{fig:fractions}
    
\end{figure*}

We define the paired fraction as the number of paired galaxies divided by the total  number of barred galaxies. In the left panel of Figure \ref{fig:fractions}, we plot the paired fraction of our sample over $\mathcal{R}$. We divide the sample into four bins in the range of  $0<\mathcal{R}<4.0$, where each bin has a width of 1. Bins of $\mathcal{R}>4.0$ were not considered due to their low sample size, leaving 288 of 300 measurements. The uncertainties are calculated using the beta distribution quantile technique as described in \citet{Cameron2011}. We see a clear overall increasing trend between the paired fraction and $\mathcal{R}$. The paired fraction starts at $5^{+6}_{-2}\%$ for the lowest values of $\mathcal{R}$ ($0<\mathcal{R}<1.0$) and increases to $31^{+15}_{-9}\%$ for the highest values of $\mathcal{R}$ ($3.0<\mathcal{R}<4.0$). Moreover, the paired fraction in the last bin is almost double the paired fraction of the entire sample, which is equal to 16\% as shown with the horizontal line in Figure \ref{fig:fractions}. This result indicates that bars with higher values of $\mathcal{R}$ (i.e. slower bars) are more likely to be in paired galaxies. In the right panel of Figure \ref{fig:fractions}, we show the paired fraction over $\Omega_p$. Here, we divide the sample into four bins between $0<\Omega_p<60$ km s$^{-1}$ kpc$^{-1}$, where each bin has a width of 15 km s$^{-1}$ kpc$^{-1}$. Bins with $\Omega_p>60$ km s$^{-1}$ kpc$^{-1}$ were excluded due to low sample size, leaving 309 of 314 measurements. No noticeable trend is found, as the paired fraction remains relatively constant near the sample mean with no significant fluctuations. This suggests that the bar pattern speed is not correlated with the probability of a galaxy being paired.

\begin{figure*}[ht!]
    \plotone{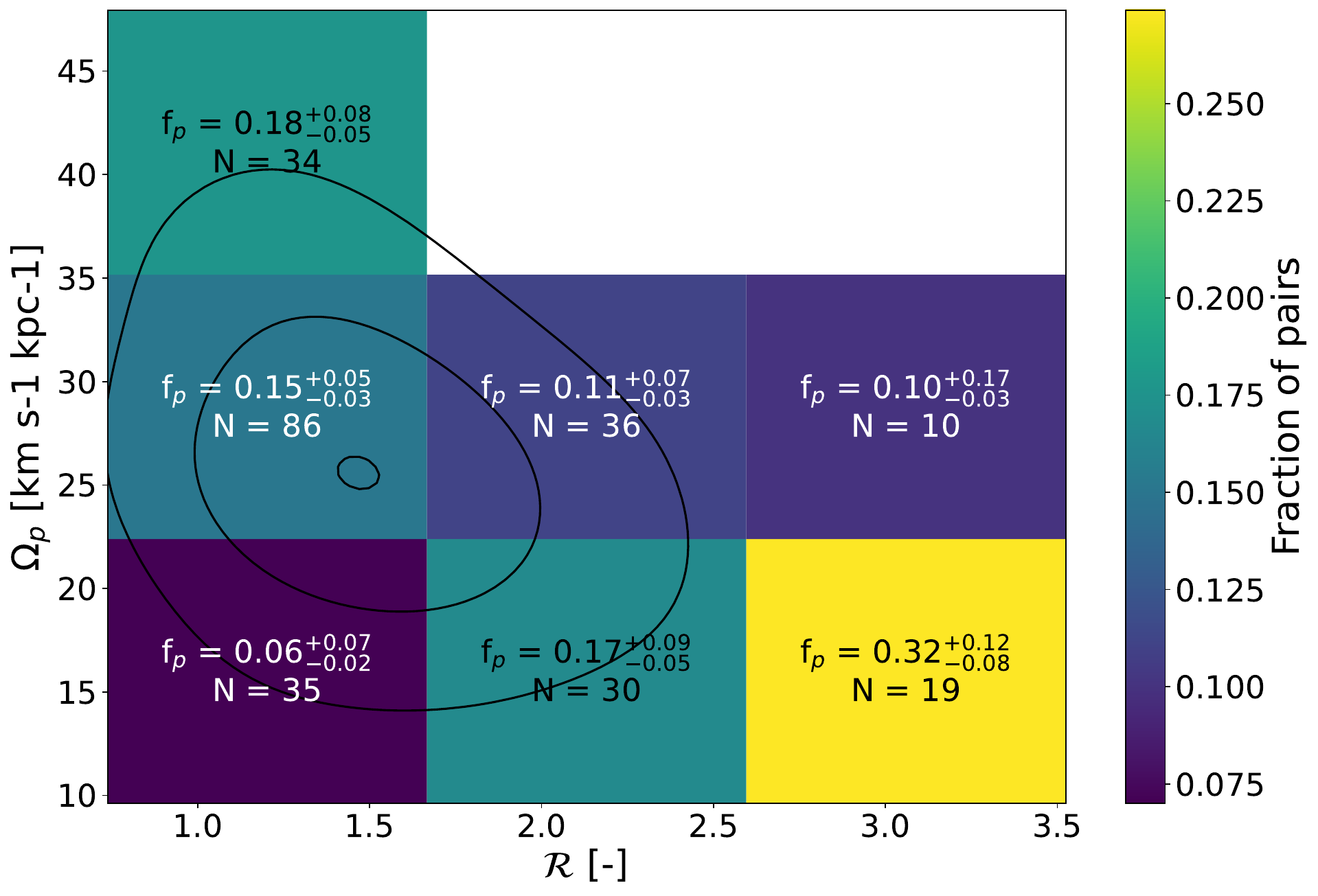}
    \caption{The paired fraction is shown in the two-dimensional ($\mathcal{R}, \Omega_p$) plane. The sample is divided into bins of $\mathcal{R}$ and $\Omega_p$, and each cell shows the corresponding bin's paired fraction, its uncertainty, and the number of measurements contained. The paired fraction is also represented by a colour scale, and the black contours show the density of the sample. The lowest 2\% and highest 5\% of both $\mathcal{R}$ and $\Omega_p$ values are removed to mitigate the effect of outliers, and only bins with a minimum of 10 galaxies are shown.}
    \label{fig:2D_frac}
\end{figure*}

We also examine the behavior of the paired fraction in the two-dimensional ($\mathcal{R}, \Omega_p$) plane as shown in Figure \ref{fig:2D_frac}. Here, we remove the lowest 2\% and highest 5\% of values in both parameters to mitigate the effect of outliers. Note that these results are largely unchanged if these thresholds are made looser. Moreover, we only include a bin if it has a minimum of 10 datapoints. We then compute the paired fraction and its uncertainty for each bin. The paired fraction attains its highest value ($32^{+12}_{-8}\%$) in the bin corresponding to the highest values of $\mathcal{R}$ and lowest values of $\Omega_p$. This indicates that bars rotating slowly with respect to both pattern speed and $\mathcal{R}$ are most likely to be found in paired galaxies in our sample. However, note that the lowest value of the paired fraction ($6^{+7}_{-2}\%$), while found in the bin of the lowest values of $\mathcal{R}$, also corresponds to the bin of the lowest values of $\Omega_p$. This indicates a trend driven largely by $\mathcal{R}$ rather than the bar pattern speed, which is consistent with the results seen in Figure \ref{fig:fractions}. In the intermediate and high $\Omega_p$ bins, we do not see any significant trends. This result shows that the trend is not necessarily steady or monotonic in this two-dimensional parameter space, but the paired fraction does attain its lowest and highest values in bins corresponding to the lowest and highest values of $\mathcal{R}$, respectively.

\begin{figure*}[ht!]
    \includegraphics[width=0.49\textwidth]{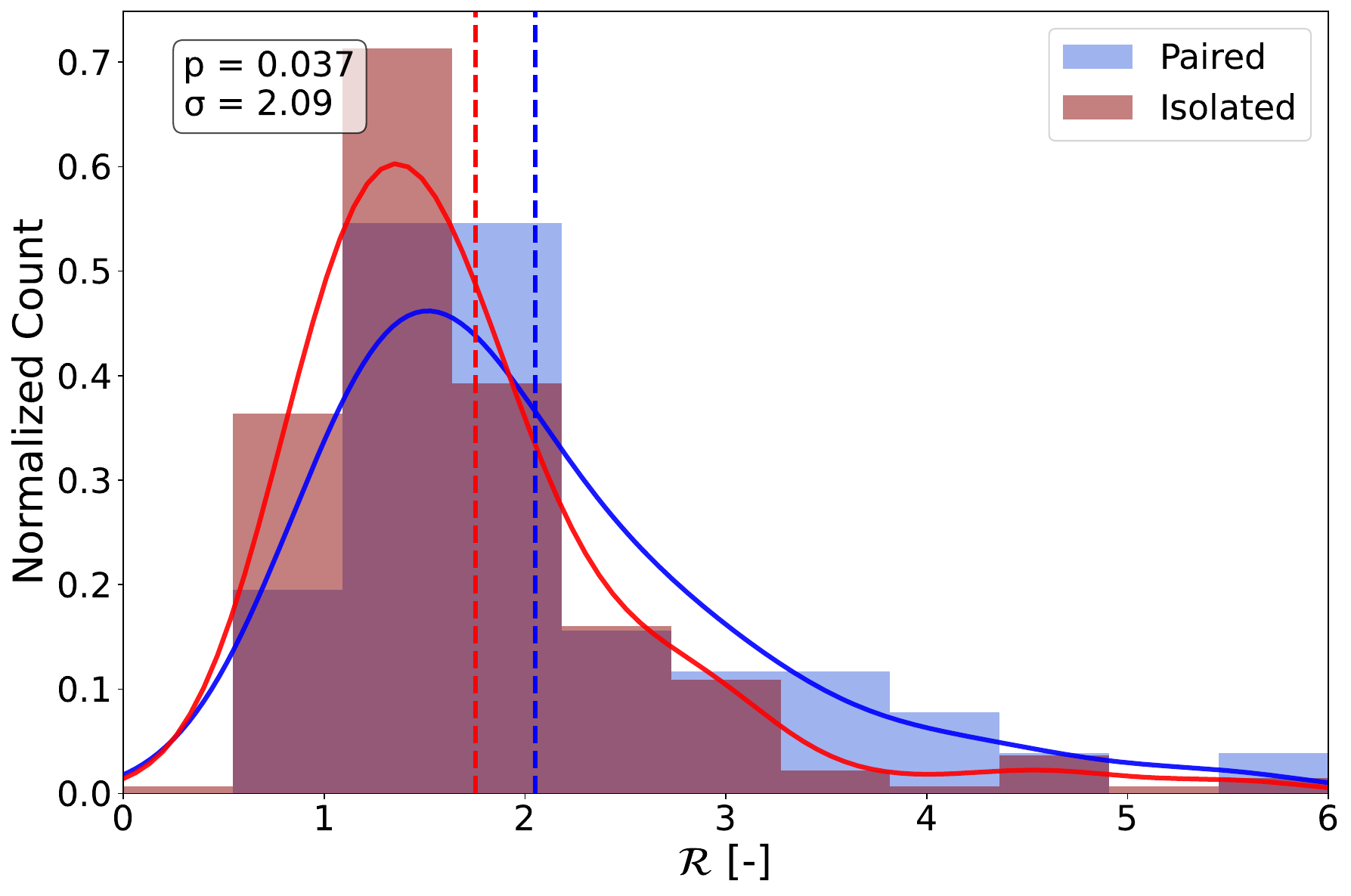}
    \includegraphics[width=0.5\textwidth]{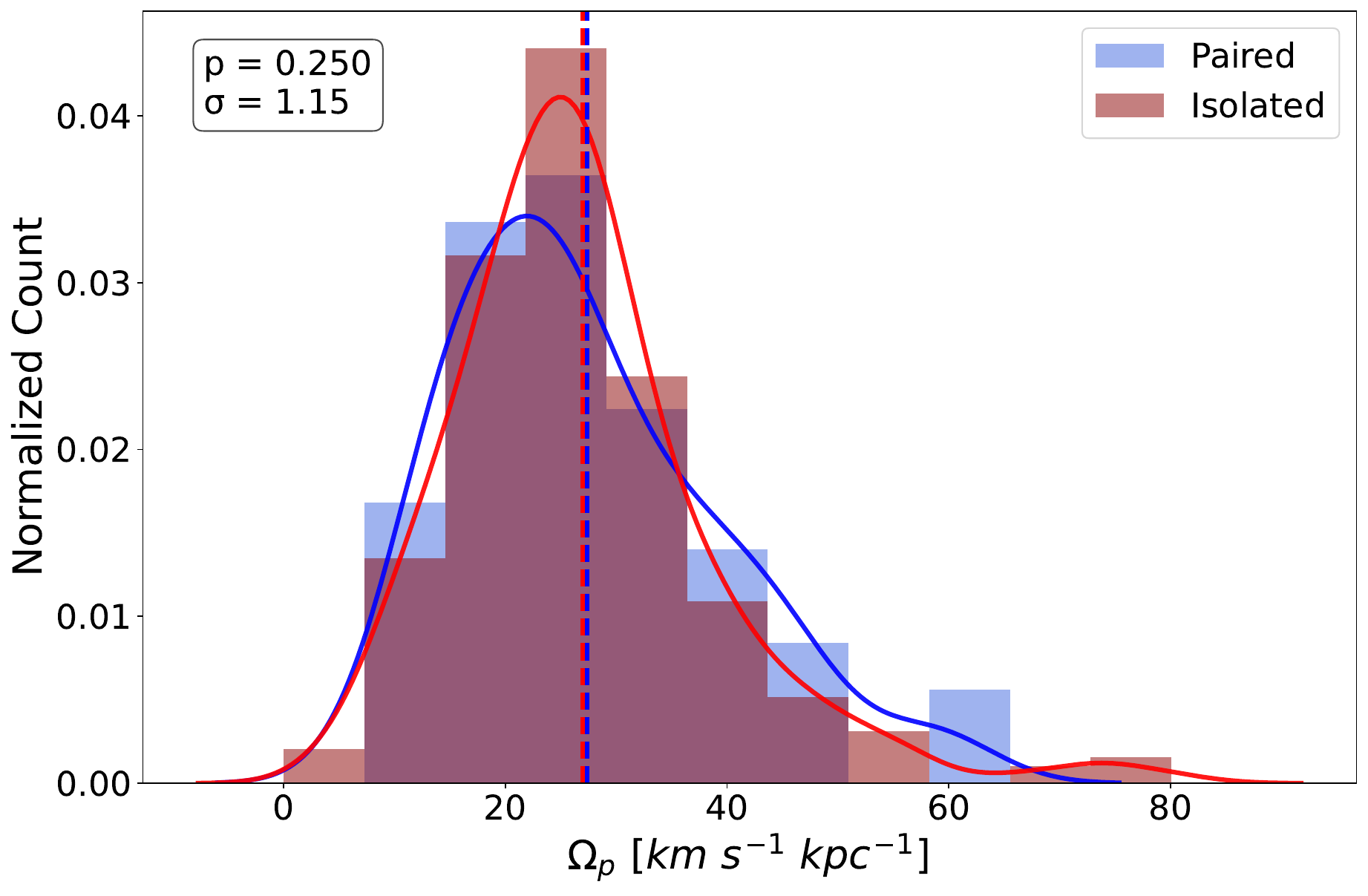}
    \caption{The distributions of $\mathcal{R}$ (left panel) and $\Omega_p$ (right panel) in paired (blue) and isolated (red) galaxies. The paired and isolated distributions of each parameter are fitted with kernel density estimates, and the mean values are indicated with the vertical dashed lines. Results of the Anderson-Darling test are also shown in the top left corner of each plot. The left panel shows the measurements of $\mathcal{R}$ for 47 paired and 253 isolated galaxies. The mean values are $\mathcal{R}=2.1\pm 0.2$ and $\mathcal{R}=1.75 \pm 0.07$, respectively. The right panel shows the measurements of $\Omega_p$ for 49 paired and 265 isolated galaxies (the full sample). The mean values are $\Omega_p = 27 \pm 2$ km s$^{-1}$ kpc$^{-1}$ and $\Omega_p=27.0 \pm 0.7$ km s$^{-1}$ kpc$^{-1}$, respectively.}
    \label{fig:R&O_hists}
\end{figure*}

In the above, we studied the behavior of the paired fraction as a function of $\mathcal{R}$ and $\Omega_p$. In Figure \ref{fig:R&O_hists}, we instead look at the entire distribution of $\mathcal{R}$ and $\Omega_p$ for paired and isolated galaxies. The left panel shows the distribution of $\mathcal{R}$ for the 253 isolated and 47 paired galaxies. The mean values of $\mathcal{R}$ in the isolated and paired galaxies are $\mathcal{R}=1.75 \pm 0.07$ and $\mathcal{R}=2.1 \pm 0.2$, respectively, although we note that both distributions have a considerable overlap. The standard error of the mean values of $\mathcal{R}$ and $\Omega_p$ throughout this work were calculated as $s/\sqrt{N}$, where $s$ is the standard deviation of the sample and $N$ is the number of galaxies in the relevant sample. To determine whether a significant difference occurs between the two populations, we perform an Anderson-Darling test \citep{ADtest}, which returns a p-value of 0.037 (2.09$\sigma$). While this does not cross the conventional $3.0\sigma$ significance threshold, it does suggest a possible difference between the two populations. Note that the relatively small sample size in each group might be a factor limiting the significance of the observed results. Moreover, the observed difference between bars in isolated and paired galaxies would be most pronounced if these bars indeed formed through internal instabilities and tidal interactions, respectively. However, this is not necessarily true for all bars in our sample. For example, bars currently in isolation may have formed through tidal interactions, and likewise bars in paired galaxies may have formed through internal instabilities. This is capable of diluting the result seen in Figure \ref{fig:R&O_hists}. Moreover, some bars in paired galaxies may have originally formed in isolation and were subsequently affected by an interaction, and we discuss this possibility in Section \ref{subsec:kin_discussion}. More generally, irrespective of the original formation mechanism, ongoing interactions involving bars in paired galaxies may influence the evolution of bar properties. Therefore, differences observed in kinematics between bars in paired and isolated galaxies may not be solely attributed to the formation mechanism. In any case, in Figure \ref{fig:R&O_hists}, the mean value of $\mathcal{R}$ for the galaxies that are paired is higher than the mean value of $\mathcal{R}$ for isolated galaxies. This is consistent with the trend observed in Figure \ref{fig:fractions}, where a positive correlation was found between $\mathcal{R}$ and the paired fraction.

The right panel of Figure \ref{fig:R&O_hists} shows the distribution of the bar pattern speed for the 265 isolated and 49 paired galaxies. The mean values in the isolated and paired galaxies are $\Omega_p=27.0 \pm 0.7$ km s$^{-1}$ kpc$^{-1}$ and $\Omega_p = 27 \pm 2$ km s$^{-1}$ kpc$^{-1}$, respectively, and the Anderson-Darling test returns a p-value of $>$0.250 ($<$1.15$\sigma$). In this case, we can conclude that there is no difference in $\Omega_p$ between paired and isolated galaxies, which supports our finding that the bar pattern speed is not correlated with the likelihood of a galaxy being paired, as seen in Figure \ref{fig:fractions}.

\subsection{Major and Minor Interactions}\label{subsec:maj/min_results}

\begin{figure*}[ht!]
    \centering
    \includegraphics[width=0.49\textwidth]{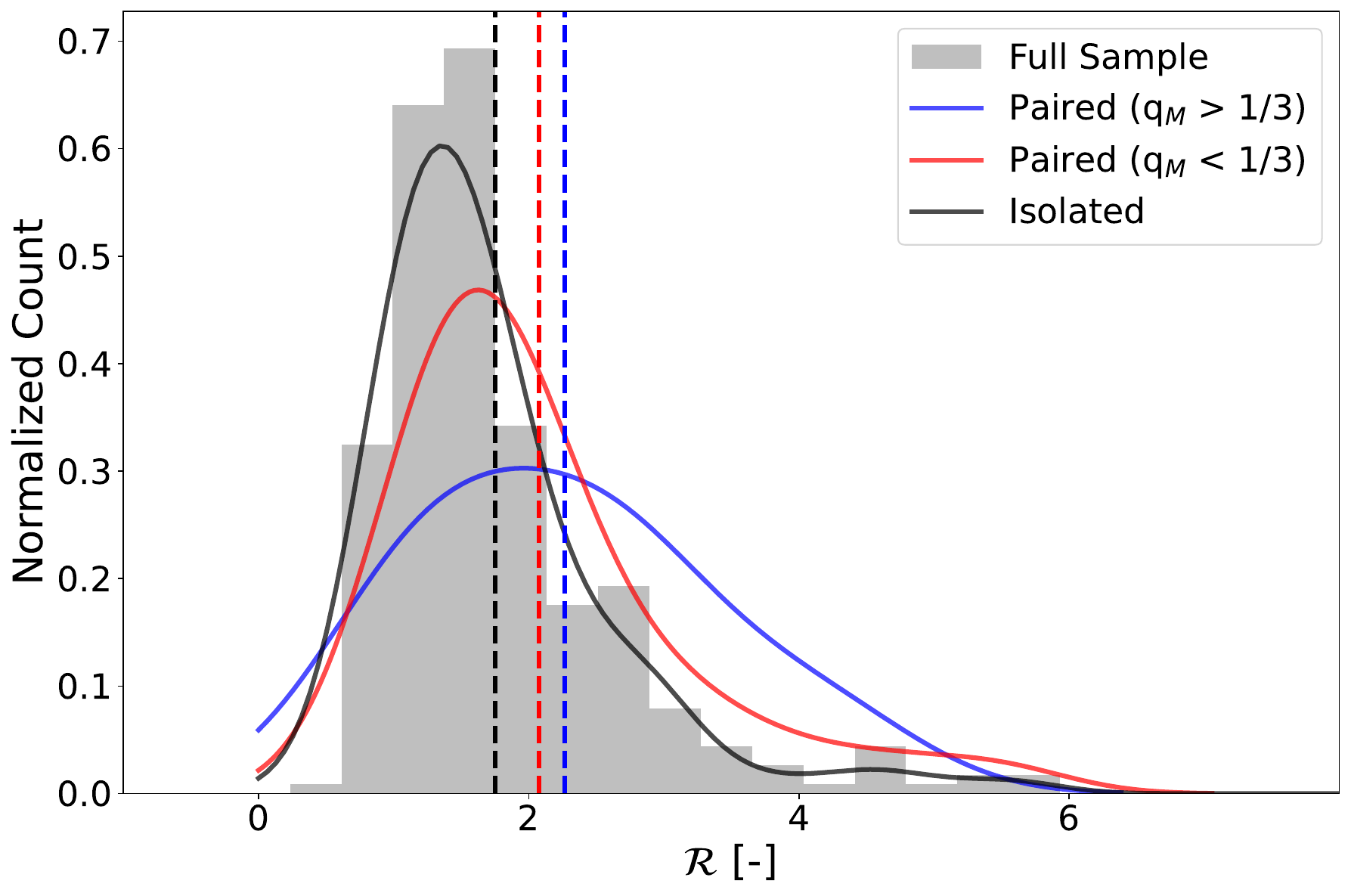}
    \includegraphics[width=0.49\textwidth]{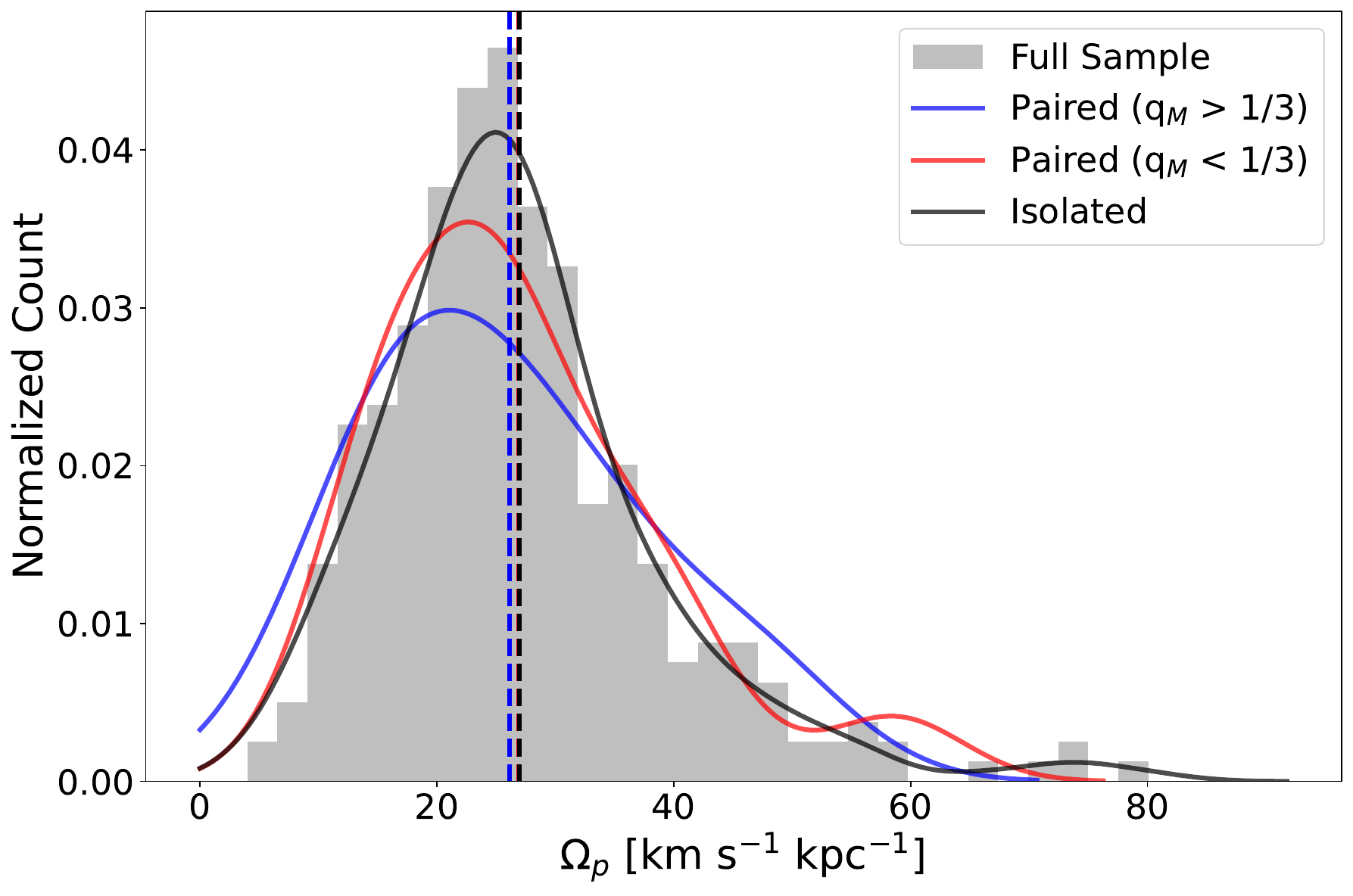}
    \includegraphics[width=0.49\textwidth]{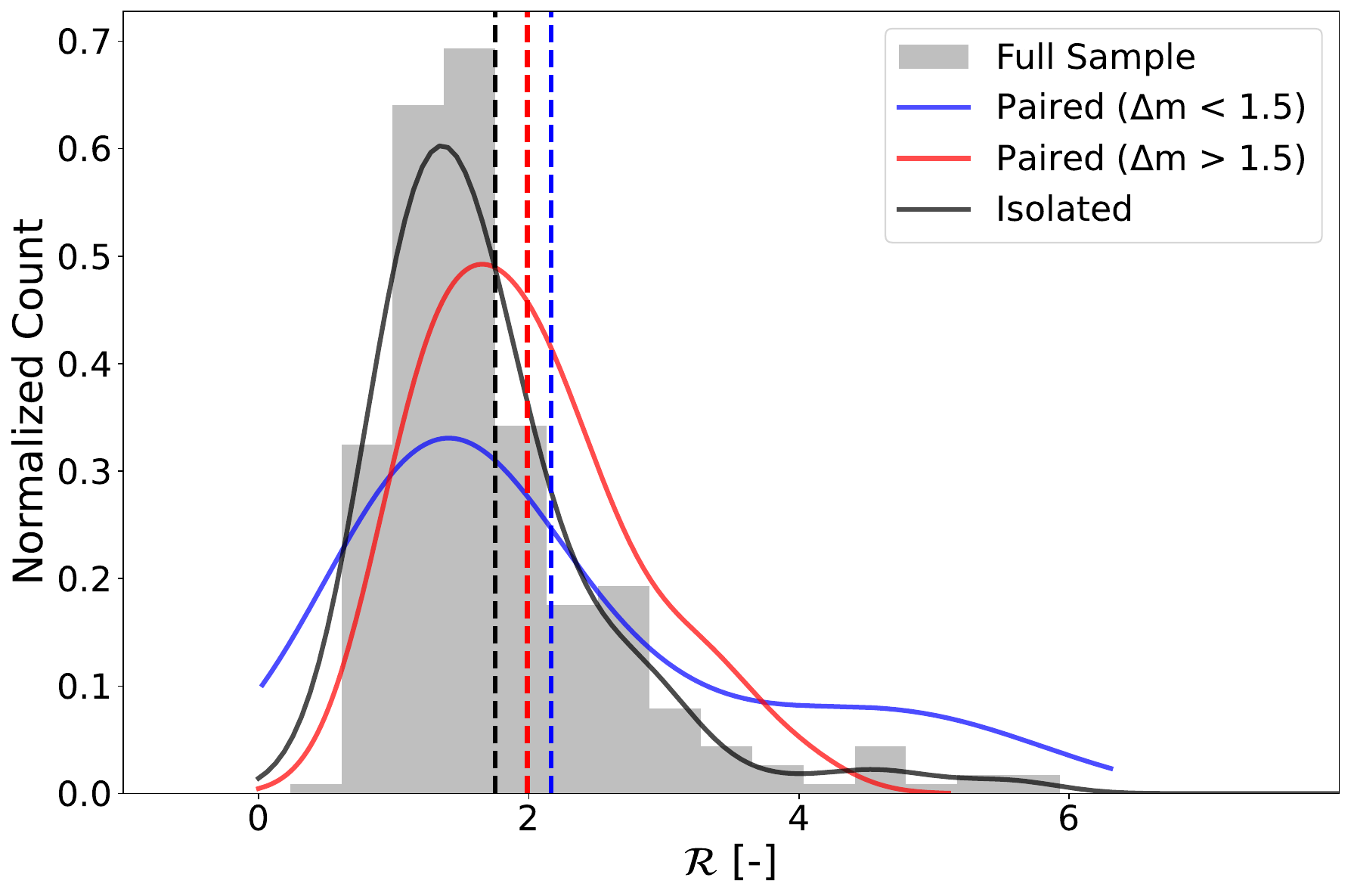}
    \includegraphics[width=0.49\textwidth]{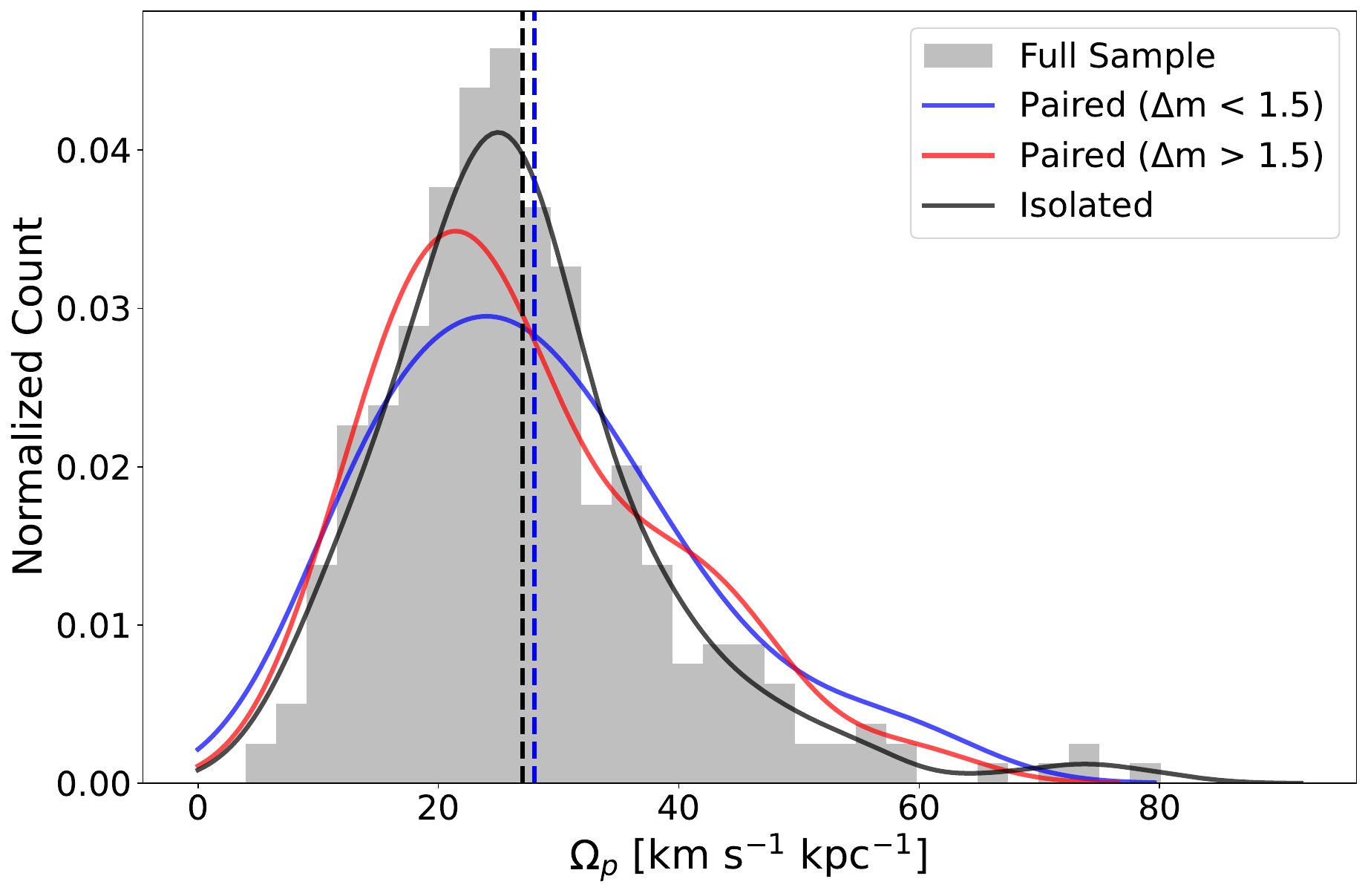}
    \caption{The distributions of $\mathcal{R}$ (left) for 300 barred galaxies and $\Omega_p$ (right) for all 314 barred galaxies. The distributions for isolated galaxies, minor interactions, and major interactions, are fitted with kernel density estimates in black, red, and blue, respectively. The vertical dashed lines represent the mean values of $\mathcal{R}$ and $\Omega_p$ for each category. The top row shows the distributions acquired using the stellar mass ratio $q_M$ to classify major and minor interactions, where a pair with $q_M > 1/3$ is considered a major interaction. The bottom row represents major and minor interactions as classified using the difference in magnitudes $\Delta m$ where $\Delta m <1.5$  represents a major interaction.}
    \label{fig:maj/min}
\end{figure*}

As described in Section \ref{subsubsec:maj/min_methods}, we computed stellar mass ratios and magnitude differences between our paired galaxies and their companions. These quantities are used to classify our paired galaxies into major interactions ($q_{\rm M}>1/3$ or $\Delta m < 1.5$) and minor interactions ($q_{\rm M}<1/3$ or $\Delta m > 1.5$). The bar pattern speed and $\mathcal{R}$ were then compared between the three following categories: isolated, major interactions, and minor interactions, using each method independently, as shown in Figure \ref{fig:maj/min}.

The top row of Figure \ref{fig:maj/min} shows a comparison of the distributions of $\mathcal{R}$ (top left panel) and $\Omega_p$ (top right panel) for isolated galaxies, pairs with major interactions, and pairs with minor interactions, as defined with the mass ratio. The mean values of $\mathcal{R}$ for isolated galaxies, pairs with minor interactions, and pairs with major interactions are $\mathcal{R}=1.75 \pm 0.07$, $\mathcal{R}=2.1 \pm 0.2$, and $\mathcal{R}=2.3 \pm 0.4$, respectively. Moreover, a slight shift towards higher $\mathcal{R}$ is seen in the distribution for major interactions. No difference is observed in terms of $\Omega_p$ for all three distributions. The mean values for isolated galaxies, pairs with minor interactions, and pairs with major interactions are $\Omega_p=27.0 \pm 0.7$ km s$^{-1}$ kpc$^{-1}$, $\Omega_p=27 \pm 2$ km s$^{-1}$ kpc$^{-1}$ and $\Omega_p=26 \pm 4$ km s$^{-1}$ kpc$^{-1}$, respectively. These results suggest that the barred galaxies in our sample with higher values of $\mathcal{R}$, as defined using the stellar mass ratio, may be associated with major interactions. However, $\Omega_p$ does not differ between major or minor interactions. Note that the sample sizes in each category are relatively low (34 minor interactions, 7 major interactions), which limits the statistical significance of the observed results. The aforementioned trends in $\mathcal{R}$, while suggestive, are therefore not statistically significant ($<3.0\sigma$). Another factor limiting the reliability of these trends is the possible uncertainty associated with the measured stellar masses in close interacting systems, as discussed in Section \ref{subsubsec:maj/min_methods}.

The bottom row of Figure \ref{fig:maj/min} shows the distributions of $\mathcal{R}$ (bottom left panel) and $\Omega_p$ (bottom right panel) for major and minor interactions defined by the magnitude difference (see Section \ref{subsubsec:maj/min_methods}). The mean values of $\mathcal{R}$ for isolated galaxies, pairs with minor interactions, and pairs with major interactions are $\mathcal{R}=1.75 \pm 0.07$, $\mathcal{R}=2.0 \pm 0.1$, $\mathcal{R}=2.2 \pm 0.4$, respectively. No difference is observed in terms of $\Omega_p$ for all three distributions. The mean values for isolated galaxies, pairs with minor interactions, and pairs with major interactions are $\Omega_p=27.0 \pm 0.7$, $\Omega_p=27 \pm 2$ km s$^{-1}$ kpc$^{-1}$, and $\Omega_p=28 \pm 3$ km s$^{-1}$ kpc$^{-1}$, respectively. The results are similar to those observed using the stellar mass ratios, where $\mathcal{R}$ values are on average the highest in galaxies with major interactions, and $\Omega_p$ values are similar in every category. Again, due to relatively low sample sizes (32 minor interactions, 17 major interactions), as well as possible uncertainties associated with the measured magnitudes (see Section \ref{subsubsec:maj/min_methods}), the observed trends in $\mathcal{R}$ are suggestive but not statistically significant. The mean values of $\mathcal{R}$ and $\Omega_p$ for paired galaxies in major and minor interactions are summarized in Table \ref{tab:majmin}.

\begin{table*}
\centering
\renewcommand{\arraystretch}{1.35}
\setlength{\tabcolsep}{28pt}

\begin{tabular}{lcc}
\hline
Sample & $\mathcal{R}$ [-]& $\Omega_p$ [km s$^{-1}$ kpc$^{-1}$]\\
\hline
Isolated Galaxies & $1.75 \pm 0.07$ & $27.0 \pm 0.7$ \\
Paired Galaxies & $2.1 \pm 0.2$ & $27 \pm 2$  \\
\hline
\multicolumn{3}{c}{Stellar Mass Ratios} \\
\hline
Minor ($q_{\rm M} < 1/3$) & $2.1 \pm 0.2$ & $27 \pm 2$ \\
Major ($q_{\rm M} > 1/3$) & $2.3 \pm 0.4$ & $26 \pm 4$ \\
\hline
\multicolumn{3}{c}{Magnitude Differences} \\
\hline
Minor ($\Delta m > 1.5$) & $2.0 \pm 0.1$ & $27 \pm 2$ \\
Major ($\Delta m < 1.5$) & $2.2 \pm 0.4$ & $28 \pm 3$ \\
\hline
\end{tabular}

\caption{The mean values of $\mathcal{R}$ and $\Omega_p$ for isolated galaxies, paired galaxies, paired galaxies in minor interactions, and paired galaxies in major interactions. The distinction of minor from major interactions is made using each of the stellar mass ratios, and the absolute $r$-band magnitude differences between paired galaxies and their companions, and the mean and uncertainty resulting from each method are shown.}
\label{tab:majmin}

\end{table*}

\section{Discussion}\label{sec:discussion}

\subsection{Bar Kinematics of Paired and Isolated Galaxies}\label{subsec:kin_discussion}

In this work, we classified barred galaxies as either paired or isolated, and compared the kinematics of bars between the two populations. The main hypothesis motivating our work is that of the potential link between the formation mechanism of a bar and its kinematics. Therefore, the observed results, and particularly the contrast between the kinematics of bars in paired and isolated galaxies, would be strongest if the bars indeed followed the corresponding formation mechanism of each population. However, it is important to note that the true formation histories of the bars in our sample are not known. Bars in paired galaxies may indeed have formed through tidal interactions, and bars in isolated galaxies may have formed through internal instabilities. On the other hand, bars in paired galaxies may not have been purely tidally induced. Instead, they may have originally formed in isolation and later affected by tidal interactions, or through a combination of internal instabilities and tidal interactions. Such bars are referred to as ``tidally affected" bars. In what follows, we discuss our results in relation to simulations that studied tidally induced bars, tidally affected bars, or both.

In the left panel of Figure \ref{fig:fractions}, we showed the paired fraction of our sample as a function of $\mathcal{R}$. From this result, we conclude that bars with higher values of $\mathcal{R}$ are more likely to be found in paired galaxies. Our results support the findings of \citet{martinez-valpuesta2016,martinez-valpuesta-2017,zheng&shen1}, whose simulations show that tidally induced bars, meaning those formed in galaxies where a bar is not able to form in isolation, are slower with respect to $\mathcal{R}$ than bars formed by internal disk instabilities in isolated galaxies. Moreover, \citet{lokas2014,lokas2016,gajda2017, gajda2018} simulated galaxies with only tidally induced bars and observed the behavior of $\mathcal{R}$. \citet{lokas2014} found that it increases from $\mathcal{R}=1.5$ to $\mathcal{R}=3.3$ after repeated interactions. \citet{gajda2017} found that all bars have values of $2.0<\mathcal{R}<4.0$, while \citet{lokas2016,gajda2018} found values of $2.0<\mathcal{R}<3.0$. While these works do not include isolated galaxies in their simulations, their obtained values of $\mathcal{R}$ for tidally induced bars place the bars in the slow category. These $\mathcal{R}$ values are also consistent with our results in Figure \ref{fig:fractions}, where higher values of the paired fraction relative to our sample are observed in the bins corresponding to $2.0<\mathcal{R}<4.0$. These works are also consistent with the distributions of $\mathcal{R}$ seen in Figure \ref{fig:R&O_hists}, where we observed a possible shift towards higher $\mathcal{R}$ in paired galaxies (p-value=0.037; 2.09$\sigma$), and a mean value of $\mathcal{R}=2.1 \pm 0.2$ for bars in paired galaxies as opposed to $\mathcal{R}=1.75 \pm 0.07$ for isolated galaxies. \citet{martinez-valpuesta2016,martinez-valpuesta-2017,zheng&shen1} also simulated bar formation in galaxies with cold and warm disks (where bars are able to form in isolation), and then subjected these bars to tidal interactions. In the simulations of \citet{martinez-valpuesta2016,martinez-valpuesta-2017}, the bar was allowed to form in isolation first, and was then subjected to an interaction, to ensure that the formation is due solely to the internal instabilities of the disk. In the simulations of \citet{zheng&shen1}, the formation of the bar, in this case, is said to be due to a combination of both internal instabilities and tidal interactions. In both cases, these bars are referred to as ``tidally affected" bars, rather than tidally induced bars. Our results also support the findings of these simulations in such cases, where it was found that tidally affected bars have higher values of $\mathcal{R}$ than bars formed in isolation, in galaxies that remain isolated. Our results are also in agreement with the observational work of \citet{puczek2026}, which found that bars in high-density environments are significantly slower with respect to $\mathcal{R}$ compared to bars in low-density environments.

In the right panel of Figure \ref{fig:fractions}, we show the paired fraction as a function of the bar pattern speed and conclude that $\Omega_p$ is not correlated with the probability of a galaxy being paired. Similarly, through the distributions of $\Omega_p$ for paired and isolated galaxies, shown in the right panel of Figure \ref{fig:R&O_hists}, we conclude that there is no difference in $\Omega_p$ between bars in paired and isolated galaxies. Our results support the findings of \citet{zheng&shen1}, which find comparable values of $\Omega_p$ in tidally affected bars and isolated bars, in galaxies where bars are able to form in isolation. However, our results are in disagreement with their simulations of kinematically hotter disks, where only tidally induced bars form, as lower values of $\Omega_p$ are found in this case. This may be due to inherent properties of the relevant galaxies, and we discuss the nuances of different disk dynamics and galaxy properties in more detail in Section \ref{subsec:properties_discussion}.

In the simulations of \citet{martinez-valpuesta2016,martinez-valpuesta-2017}, bars formed in isolation have lower values of $\Omega_p$ after an interaction than those which remain isolated. However, the difference is only around $\Omega_p = 2$ km s$^{-1}$ kpc$^{-1}$ (see Figure 1c of \citealp{martinez-valpuesta2016}). This difference is much smaller than the observational uncertainties in this work, where the median uncertainty in the pattern speed is $\pm 5$ km s$^{-1}$ kpc$^{-1}$. Therefore, such small differences can not be resolved with the sample used in this work. However, our results are again in disagreement with the simulations of \citet{martinez-valpuesta2016,martinez-valpuesta-2017} in the case of disks where only tidally induced bars form. In this case, these simulations find much lower values of $\Omega_p$. Furthermore, \citet{lokas2014} simulated repeated interactions of a barred dwarf galaxy and found that the pattern speed experiences sharp drops at the pericentres, but almost fully recovers after each interaction. Similarly, the simulations of \citet{jimenez-arranz2025} found that in the Large Magellanic Cloud (LMC), a barred dwarf galaxy, the bar pattern speed slows down significantly after an interaction, leading to a complete halt or very slow rotation of the bar. After some time, the pattern speed recovers its initial value before the interaction. However, note that these are results of simulations involving bars in dwarf galaxies whose stellar mass is far smaller than the median stellar mass of the galaxies in our sample, which could be a factor affecting the behavior of the bar kinematics after an interaction.

\subsection{Major and Minor Interactions}\label{subsec:maj/min_discussion}

The results of Figure \ref{fig:maj/min} suggest that bars with higher values of $\mathcal{R}$ may be associated with major interactions. However, there is considerable overlap between the distributions of $\mathcal{R}$ for major and minor interactions. Moreover, note that our sample size of major interactions is particularly small, and the observed trend is not statistically significant. This result may also be subject to additional uncertainties due to potential difficulties in measuring the magnitudes and stellar masses of galaxies in interacting systems, as discussed in Section \ref{subsubsec:maj/min_methods}. \citet{zheng&shen1} defined strong interactions as pairs with a mass ratio $q_{\rm M}=1$, moderate interactions as pairs with $q_{\rm M} = 1/3$, and weak interactions as pairs with $q_{\rm M}=1/10$. With this distinction, they found no significant differences in values of $\mathcal{R}$, in tidally induced bars, between strong, moderate, and weak interactions. However, in the work of \citet{martinez-valpuesta2016,martinez-valpuesta-2017}, where tidally induced bars had higher values of $\mathcal{R}$, all simulated interactions have a mass ratio $q_{\rm M}=1$. Therefore, it is worth questioning whether the differences in $\mathcal{R}$ in those works are more pronounced due to the 1:1 nature of the interactions, not the formation mechanism alone. Moreover, we found that the bar pattern speed remains the same for isolated galaxies, pairs with major interactions, and pairs with minor interactions. This is consistent with the simulations of \citet{zheng&shen1}, where no differences were found in values of $\Omega_p$, in tidally induced bars, between strong ($q_{\rm M}=1$), moderate ($q_{\rm M} = 1/3$), and weak ($q_{\rm M}=1/10$) interactions.

\subsection{The Effect of Galaxy and Interaction Properties on Bar Kinematics}\label{subsec:properties_discussion}

Our results are generally in agreement with those of several simulations, as explained in Section \ref{subsec:kin_discussion}. However, there are unique cases with special considerations of galaxy properties or interaction types that lead to contradicting results. For example, \citet{lokas2018} simulated interactions of varying strength and found that interactions with lower relative velocities, considered to be stronger interactions, slow down the bar with respect to $\Omega_p$. \citet{lokas2016} also simulated different types of orbits of interacting galaxies and found that tighter orbits lead to lower values of $\Omega_p$. Another example is found in \citet{zheng&shen1}, where only prograde orbits led to considerably lower values of $\Omega_p$ in tidally affected bars compared to bars formed in isolation, in galaxies where bars are able to form in isolation. However, in perpendicular and retrograde orbits, the values of $\Omega_p$ were largely similar to those of bars formed in isolation. 

Some works interpret the difference in kinematics between tidally induced bars and bars formed in isolation as a result of inherent galaxy properties and not the formation mechanism. \citet{zheng&shen1} examined bar kinematics in three disk models: cold and warm disks (where bars are able to form both in isolation and through tidal interactions), and hot disks (where only tidally induced bars can form). When comparing tidally affected bars and bars formed in isolation in the cold and warm disks, relatively similar values of $\mathcal{R}$ were seen. Only in the tidally induced bars in hot disks were higher values of $\mathcal{R}$ consistently seen across all interaction strengths. Similar results were seen for $\Omega_p$, where lower values of $\Omega_p$ were seen in the kinematically hot disks. Therefore, \citet{zheng&shen1} argue that differences in kinematics between tidally induced bars and bars formed in isolation are due to internal properties of the host galaxy of the bar, not the different formation mechanisms. They also note that if perturbations promote or delay bar formation, as suggested by \citet{moetazedian2017,pettitt2018,zheng&shen2}, tidally induced bars may not be at the same evolutionary stages as bars formed in isolation, and this may reflect a difference in kinematics. All of the results discussed in this subsection serve to show that differences in bar kinematics may not be solely attributed to different formation mechanisms. Instead, the kinematics are affected by other factors, such as the dynamical state of the disk and the properties of the interaction. However, it is difficult to disentangle such factors with observations. This is capable of diluting the results presented in this work, and may explain the large overlap between the values of $\mathcal{R}$ in the left panel of Figure \ref{fig:R&O_hists}.

As described in Section \ref{subsubsec:kin_methods}, the bar radius was measured differently in each of the four works from which we obtain kinematic measurements of the bars. There is no method of measuring the bar radius that is definitively the best, as discussed by \citet{Guo,garma-oehmichen2022}. The difficulty is mainly due to the fact that bars do not have a well-defined physical boundary. Moreover, host galaxy morphology can affect the suitability of a certain method of measuring the bar radius. A galaxy's orientation, as well as structures (spirals, rings, etc.), can affect where a certain method defines the bar's end. For example, \citet{garma-oehmichen2020,garma-oehmichen2022} use the isophotal ellipticity profile and the disk PA profile to measure the bar radius. In some galaxies, the maximum ellipticity is in fact produced by open spiral arms that follow the bar, not the bar itself \citep{garma-oehmichen2022}. It is for these reasons that \citet{Guo} took the average of three $R_{\rm bar}$ measurements using the three most widely used methods: the isophotal ellipticity profile, the disk PA profile, and the Fourier decomposition. For more detailed discussions on the specific biases introduced by these methods, see \citet{Guo,garma-oehmichen2022}. This is important since the measurement of the bar radius has a direct effect on the measurement of $\mathcal{R}=R_{\rm CR}/R_{\rm bar}$. If the bar radius is underestimated, then $\mathcal{R}$ would be overestimated, and vice versa. Since our work adopts measurements from studies that used different methods of measuring $R_{\rm bar}$, this may introduce additional scatter into the $\mathcal{R}$ distribution. A similar point can be made regarding the circular velocity $V_c$, which is used in \citet{Guo} to find the corotation radius as $R_{\rm CR}=V_c/\Omega_p$. In this calculation, $V_c$ is the circular velocity in the flat part of the rotation curve. It is not necessarily the case that the corotation radius lies in this part of the curve, and this assumption may lead to incorrect estimates of $R_{\rm CR}$, and in turn, $\mathcal{R}$ \citep{geron2023}. To avoid this assumption, \citet{garma-oehmichen2020,garma-oehmichen2022, geron2023} instead chose to model the actual rotation curves and use them to obtain $R_{\rm CR}$, as explained in Section \ref{subsubsec:kin_methods}.

Aside from the intricacies of disk kinematics, other factors contribute to the difficulty of observing the slowdown of bars with respect to $\Omega_p$ in paired galaxies. For one, simulations show that the bar pattern speed naturally decreases over time \citep{athanassoula2003,sellwood2006}. This can easily dilute the difference seen in $\Omega_p$ between paired and isolated galaxies if the observed bars have formed at different times. In this work, the ages of the observed bars are unknown. Therefore, when a bar which is slow with respect to $\Omega_p$ is observed, it is unclear whether the bar has slowed down over time, or whether it was formed with an initially low pattern speed. Another factor affecting the evolution of the bar pattern speed is the gas content of the host galaxy. Simulations have shown that pattern speeds can decelerate rapidly in gas-poor galaxies, while they remain relatively constant or decrease much more gradually in gas-rich galaxies \citep{berentzen2007,villa-vargas2010}. For these reasons, although the local environment of a galaxy may be a good gauge for the changes in the bar rotation rate $\mathcal{R}$, as shown in this work, perhaps another approach is needed to study the behavior of the pattern speed. On the other hand, the slowdown of $\Omega_p$ over time has been tested using investigations of the mass of the host galaxy. Simulations by \citet{kataria2019} show that the pattern speed decreases at a faster rate in galaxies with more massive galactic bulges. The simulations of \citet{semczuk2024} also show that bars with an initially high pattern speed and a subsequent slowdown are more likely found in massive galaxies, while lower mass galaxies host bars that remain at a relatively low pattern speed. \citet{geron2026} tested this prediction observationally and found that the pattern speed significantly decreases with both the stellar mass and the total dynamical mass of galaxies. In short, differences in the bar pattern speed are due to several factors which can be difficult to disentangle observationally using only the local environment of host galaxies.

\section{Conclusion}\label{sec:conc}

Examining the kinematics of a bar in relation to the local environment of its host galaxy provides meaningful insight into possible correlations between the formation history of a bar and its kinematic behavior. Simulations suggest that tidally induced bars rotate slower than those formed in isolation, and this work provides an observational test of this prediction. We used measurements of the bar rotation rate ($\mathcal{R}$) and the bar pattern speed ($\Omega_p$) of 314 barred galaxies from four studies \citep{Guo,garma-oehmichen2020,garma-oehmichen2022,geron2023} which obtained these measurements using IFU data from the MaNGA survey. We examined the local environment of each barred galaxy, here defined as nearby potential companion galaxies, using the GZ DESI \citep{GZDESI} catalogue. We then classified each galaxy as either paired or isolated, and we proceeded to analyze the kinematics of bars in paired and isolated galaxies. Our main results are as follows:

\begin{enumerate}
    \item We conclude that bars in our sample with higher values of $\mathcal{R}$ are more likely to be found in paired galaxies. This was determined after a clear overall increasing trend was observed between $\mathcal{R}$ and the paired fraction in our sample. The paired fraction starts at $5^{+6}_{-2}\%$ for the lowest values of $\mathcal{R}$ ($0<\mathcal{R}<1.0$) and increases to $31^{+15}_{-9}\%$ for the highest values of $\mathcal{R}$ ($3.0<\mathcal{R}<4.0$). On the other hand, no correlation was observed between $\Omega_p$ and the paired fraction. We therefore conclude that $\Omega_p$ is not correlated with the probability of a galaxy being paired. We also examine the behavior of the paired fraction in the two-dimensional ($\mathcal{R}$, $\Omega_p$) space. We conclude that trends involving the paired fraction are driven largely by $\mathcal{R}$, which reinforces the previous results.
    \item We have found a possible difference in the distribution of $\mathcal{R}$ between paired galaxies and isolated galaxies. The mean values are $\mathcal{R}=1.75\pm 0.07$ and $\mathcal{R}=2.1 \pm 0.2$ in isolated and paired galaxies, respectively. However, there is considerable overlap between the distributions of $\mathcal{R}$ and the resulting difference is not statistically significant (p-value=0.037; 2.09$\sigma$). Future work would benefit from an increased sample size to clarify whether a statistically significant difference occurs in $\mathcal{R}$. While we find a possible difference for $\mathcal{R}$, we conclude that there is no difference in the distribution of $\Omega_p$ between paired and isolated galaxies (p-value$>$0.250; $<$1.15$\sigma$). The mean values in isolated and paired galaxies are $\Omega_p=27.0\pm 0.7$ km s$^{-1}$ kpc$^{-1}$ and $\Omega_p=27\pm2$ km s$^{-1}$ kpc$^{-1}$, respectively.
    \item Our results suggest that bars with higher values of $\mathcal{R}$ may be associated with major interactions. We define major and minor interactions using the stellar mass ratio $q_{\rm M}$ and the difference in magnitudes $\Delta m$. However, our sample size is considerably reduced after distinguishing between major and minor interactions, and thus these results are not statistically significant. This result may also be subject to uncertainties as a result of difficulties in measuring the magnitudes and stellar masses of galaxies in interacting systems. On the other hand, we find no difference in the bar pattern speed between major and minor interactions.

\end{enumerate}

Based on these conclusions, our findings are in agreement with simulations showing that tidally induced or tidally affected bars tend to rotate slowly with respect to $\mathcal{R}$. However, our results disagree with simulations showing lower values of the bar pattern speed in tidally induced or tidally affected bars, while noting the argument that such differences are due to internal properties of the galaxies involved. Future work with an increased sample size is needed to clarify whether the difference in the distribution of $\mathcal{R}$ between paired and isolated galaxies is statistically significant, and to further establish the lack of a difference in the bar pattern speed. Future work would also benefit from an investigation of changes in the corotation radius $R_{\rm CR}$ and the bar radius $R_{\rm bar}$ in paired and isolated galaxies, as this would help better understand the behavior of the bar rotation rate, as it is defined as $\mathcal{R}=R_{\rm CR} / R_{\rm bar}$.

\begin{acknowledgements}

TG is a Canadian Rubin Fellow at the Dunlap Institute. The Dunlap Institute is funded through an endowment established by the David Dunlap family and the University of Toronto. TG is also supported through the LSST-DA Catalyst Fellowship; this publication was thus made possible through the support of grant 62192 from the John Templeton Foundation to LSST-DA.

ILG has received the support from the Czech Science Foundation Junior Star grant no. GM24-10599M.
    
\end{acknowledgements}

\software{\texttt{Astropy} \citep{astropy2013,astropy2018,astropy2022}, \texttt{Matplotlib} \citep{matplotlib2007}, \texttt{NumPy} \citep{harris2020array}, \texttt{SciPy} \citep{2020SciPy-NMeth}, \texttt{Seaborn} \citep{seaborn2021}, \texttt{Pandas} \citep{pandas2011}}

\bibliographystyle{aasjournalv7}
\bibliography{bib}

\end{document}